\documentstyle[twoside,fleqn,espcrc2]{article}

\newcommand{\be}{\begin{equation}}
\newcommand{\ee}{\end{equation}}
\newcommand{\bea}{\begin{eqnarray}}
\newcommand{\eea}{\end{eqnarray}}
\newcommand{\bc}{\begin{center}}
\newcommand{\ec}{\end{center}}
\newcommand{\bi}{\begin{itemize}}
\newcommand{\ei}{\end{itemize}}

\title{
\vspace{-2.5cm}
       {\normalsize \hfill  DESY 03-040}   \\ [0.8cm]
\bf Accelerating the Hybrid Monte Carlo algorithm}
\author{A. Ali Khan\address[HU]{Institut f\"ur Physik, Humboldt-Universit\"at 
zu Berlin,  12489 Berlin, Germany}, 
T. Bakeyev\address[DU]{Joint Institute for Nuclear Research, 141980 Dubna,
Russia},
M. G\"ockeler\address[LU]{Institut f\"ur Theoretische Physik, Universit\"at
Leipzig, 04109 Leipzig, Germany}\address[RU]{Institut f\"ur 
Theoretische Physik, Universit\"at
Regensburg, 93040 Regensburg, Germany},
R. Horsley\address[EU]{School of Physics, The University of 
Edinburgh, Edinburgh EH9 3JZ, UK}, 
D. Pleiter\address[NIC]{John von Neumann-Institut f\"ur Computing NIC, 
15738 Zeuthen, Germany}, 
P. Rakow\addressmark[RU]\address[LI]{Theoretical 
Physics Division, Department of 
Mathematical Sciences, University of Liverpool, 
Liverpool L69 3BX, UK}, 
A. Sch\"afer\addressmark[RU],
G. Schierholz\addressmark[NIC]\address[DESY]{Deutsches 
Elektronen-Synchrotron  DESY, 22603 Hamburg, Germany} and
H. St\"uben\address{Konrad-Zuse-Zentrum f\"ur Informationstechnik
Berlin, 14195 Berlin, Germany} \\ 
QCDSF Collaboration
}

\vspace{0.3cm}

\begin{document}

\begin{abstract}
An algorithm for separating the high- and low-frequency molecular
dynamics modes in Hybrid Monte Carlo (HMC) simulations of gauge
theories with dynamical fermions is presented. The separation is based on 
splitting the pseudo-fermion action into two parts, as was initially proposed 
by Hasenbusch. 
%in \cite{Hasenbusch:2001ne}
%hep-lat/0107019 
We propose to introduce
different evolution time-scales for each part.
We test our proposal in realistic simulations of two-flavor 
$O(a)$ improved Wilson fermions.
%QCD with clover-improvement.
A speed-up of more than a factor of three compared to the standard
HMC algorithm is observed 
%for a
%$16^3\times 32$ lattice with parameters
%$\beta=5.29, \kappa=0.1355, c_{sw}=1.9192$. 
in a typical run.

%\vspace{1pc}
\end{abstract}

\maketitle

%\pacs{PACS numbers: 11.10.Wx, 11.15.Ha, 11.30Rd, 12.38.Gc}

\section{Introduction.} 

The numerical simulations of QCD with Wilson fermions using the HMC
algorithm \cite{Duane:de} provide a significant 
challenge as the quark masses
become smaller. Any improvement to make the simulations faster will help 
to increase the overlap between the domain of chiral perturbation theory and 
lattice QCD, allowing for an extrapolation to physical quark masses
\cite{Bernard:2002yk}.

In the standard implementation of HMC one introduces pseudo-fermion fields to 
take into account the contribution of the fermion determinant. 
As the quark mass becomes lighter, 
the force induced by pseudo-fermions produces increasingly large
high-frequency fluctuations. One is therefore forced to decrease the step-size
of the integration scheme to keep a constant acceptance rate.
This problem is addressed in the literature as
``ultra-violet slowing-down'' \cite{Borici:2002zu}. 

A possible solution of this problem is the introduction 
of multiple time-scales 
for different parts of the action in performing the discrete
integration of molecular dynamics (MD) 
equations of motion in the fictitious time. This approach was 
initially advocated in Ref.~\cite{Sexton:nu}, where the authors 
proposed to introduce different time-scales for the Yang-Mills term and 
the pseudo-fermion action.
Since the number of arithmetic operations required to evaluate the 
pure gauge force is much smaller than the one needed 
to evaluate the pseudo-fermion force, 
one can keep a smaller step-size for the pure gauge part of the action.
However for light fermions, the highest frequency fluctuations
belong mainly to the pseudo-fermion action, so the approach of 
Ref.~\cite{Sexton:nu} gives only a moderate improvement in that case. 

In Ref.~\cite{Peardon:2002wb} it was suggested that a 
multiple time-scale scheme is efficient only 
if one can split the action 
\be S =  S_{UV} + S_{IR} \label{UVIR}\ee 
in a way to satisfy the following 
two criteria simultaneously:
\bi
\item the force term generated by $ S_{UV} $ 
is cheap to compute compared to $S_{IR}$;
\item  the splitting (\ref{UVIR}) mainly captures the 
high-frequency modes of the 
system in $ S_{UV} $ and the low-frequency modes in $S_{IR}$.
\ei
If these criteria are met, one can keep a relatively large
step-size for the ``infra-red'' part of the action $S_{IR}$ (which generates
the computationally more expensive force term)
and relax the step-size for the ``ultra-violet'' part  
$ S_{UV} $, while the quark mass is becoming smaller.

Ref.~\cite{Peardon:2002wb} proposed to use a low-order polynomial 
approximation for mimicking the high-frequency modes of the
pseudo-fermion action.
The algorithm was tested for the 2D Schwinger model with Wilson fermions, 
producing a substantial speed-up in comparison with the standard HMC 
implementation.

In the present work we are testing a different approach. In 
Ref.~\cite{Hasenbusch:2001ne} it was proposed to split the pseudo-fermion action
into two parts, partially separating the small and large eigenvalues
of the Dirac matrix. This splitting reduces 
the condition number of the fermion matrix, allowing for a larger step-size.
In Refs.~\cite{Hasenbusch:2001xh,Hasenbusch:2002wi,Hasenbusch:2002ai} 
the method was developed and successfully applied to
four-dimensional lattice QCD with two flavors of Wilson fermions.
We propose to further improve this method by putting the two contributions
of the pseudo-fermion action from Ref.~\cite{Hasenbusch:2001ne} on different 
time-scales of the integration scheme. 

Our proposal was already considered by Hasenbusch while preparing 
Ref.~\cite{Hasenbusch:2001ne},
but he found no advantage (see Ref.~\cite{Hasenbusch:2002ai}).
Since this statement
referred to tests within the two-dimensional Schwinger model, 
we have decided to repeat the tests for lattice QCD to see if one 
can profit from the ``multiple time-scales idea'' there.
We have found that for two-flavor $O(a)$ improved Wilson
fermions 
the introduction of different 
time-scales for the 
splitting chosen as in Ref.~\cite{Hasenbusch:2001ne},
indeed gives some speed-up compared to the case where the time-scale is 
the same for both parts. 

In all our runs we have used the educated initial guess (chronological 
inversion method) proposed in Ref.~\cite{Brower:1995vx}. This method estimates 
the trial solution
for the matrix inversion as a linear superposition of a 
sequence of solutions in the recent past while performing 
the integration along the MD 
trajectory. (It was always checked that the accuracy of inversion 
was sufficient
to make the solution effectively exact and keep the algorithm reversible.)
%The less is the step-size of MD integration scheme, the more 
%chronological inversion 
%method enjoys his efficiency. The installation of the new improvements, 
The smaller the step-size of the MD integration scheme, the more
efficient the chronological inversion method is. Hence the new improvements,
which increase
the effective step-size, may seem less efficient because the chronological 
guess becomes worse.
%Therefore, the improvement factor, which we observed 
%from applying the split of the pseudo-fermion 
%action and the introduction of multiple time-scales,
%is less pronounced 
Therefore, the improvement factor which we gained from splitting
the pseudo-fermion action and introducing multiple time-scales
is less pronounced 
than it would be if we had not used the chronological inversion. 
(We did not switch off the chronological guess because our 
tests were part of a production run.) One can hope 
to gain relatively 
more from the method tested in this paper if one is not using the 
educated guess. Nevertheless, it would be naive to say that the chronological
inversion makes things worse. 
As the quark mass decreases, the step-sizes for each time-scale inevitably 
decrease, so the chronological inversion method 
will realize its potential after all, 
making the increase of computer power requirements smoother.

The paper is organized as follows. In section~\ref{smodel}  
we recall lattice QCD with $O(a)$ improved
\cite{Sheikholeslami:1985ij}, even-odd preconditioned
\cite{Jansen:1996yt,Luo:1996tx}
Wilson fermions. We also briefly describe the splitting of the pseudo-fermion 
action proposed in Ref.~\cite{Hasenbusch:2001ne}.
In section~\ref{smult} we discuss the multiple time-scales integration 
scheme
in the HMC algorithm and specify the splitting of the action 
(the choice of time-scales) for the
models which we are going to simulate. In section~\ref{ssimdet}
we give the details of our simulations and present the results.
Conclusions follow in section~\ref{sconc}.

\section{The model}\label{smodel}

We test our proposal by simulating two-flavor QCD with 
$O(a)$ improved \cite{Sheikholeslami:1985ij},
even-odd preconditioned  Wilson fermions. 
One of the possible effective actions for a
standard HMC simulation of this theory
is given in Refs.~\cite{Jansen:1996yt,Luo:1996tx}:
\be S_0[U,\phi^{\dag},\phi] = 
S_G[U] + S_{det}[U] + \phi^\dag (Q^\dag Q)^{-1} \phi\; . \label{act0}\ee
Here $S_G[U]$ is the usual Wilson plaquette action,
$\phi^\dag,\phi$ are pseudo-fermion fields, 
\be S_{det}[U] = -2 \mbox{Tr} \log(1+T_{oo})\; , \ee
\be Q= ({\bf 1} + T)_{ee} - M_{eo} ({\bf 1} + T)_{oo}^{-1} M_{oe}\; ,
\ee
$T_{ee}(T_{oo})$ is the clover matrix (diagonal in coordinate space)
on the even (odd) sites
\be (T)_{a\alpha,b\beta}(x) =
{i \over 2}c_{sw}\kappa\sigma^{\alpha\beta}_{\mu\nu}
{\cal F}^{ab}_{\mu\nu}(x)\; ,\ee
and the off-diagonal parts $M_{eo}$ and $M_{oe}$, which connect the 
even with odd
and odd with even sites, respectively, are the usual Wilson hopping matrices
(see Ref.~\cite{Jansen:1996yt} for further details). 

According to Ref.~\cite{Hasenbusch:2001ne} we start to
modify the action (\ref{act0}) by introducing 
other pseudo-fermion fields $\chi^\dag,\chi$:
\bea && S_1[U,\phi^{\dag},\phi] = 
S_G[U] + S_{det}[U] + \nonumber \\ &&
\phi^\dag W(Q^\dag Q)^{-1} W^\dag \phi
+ \chi^\dag (W^\dag W)^{-1} \chi 
\; , \label{act1}\eea
where $W$ is some auxiliary matrix. 
The idea of this modification is that 
$W$, as well as $QW^{-1}$, have smaller condition numbers than
the original matrix $Q$. This reduces the fluctuations of 
the HMC Hamiltonian at the end of the MD trajectory, 
allowing for a larger step-size in the HMC 
simulation at the same acceptance rate.

We consider here only the following choice
of the matrix $W$ \cite{Hasenbusch:2002ai}:
\be W = Q + \rho  \; ,\label{operrho}\ee
which depends on one real parameter 
$\rho$.\footnote{Another possibility, discussed in 
Refs.~\cite{Hasenbusch:2001xh,Hasenbusch:2002wi,Hasenbusch:2002ai}, is
\begin{displaymath} 
W = Q + i\gamma_5 \rho  \; .
\end{displaymath} 
We did not consider this, although the generalization
of our approach to this choice of the auxiliary matrix
is straightforward.} Up to the multiplication by a constant factor this
matrix is equivalent to the original proposal of Ref.~\cite{Hasenbusch:2001ne}.

The modification of the pseudo-fermion action (\ref{act1}) can be easily 
implemented, if a standard HMC program is already available (see 
Refs.~\cite{Hasenbusch:2001ne,Hasenbusch:2002ai} for details).

\section{Multiple time-scales}\label{smult}

The introduction of multiple time-scales for different segments of the action
in the HMC method
was initially proposed by the authors of Ref.~\cite{Sexton:nu}.
Following their idea, one constructs a reversible 
integrator $V_M\left(\tau\right)$ 
for the action (\ref{UVIR}) by
\bea && V_M\left(\tau\right) = V_{IR}\left({\tau\over 2}\right) \cdot \nonumber \\ && 
\left[V_{UV}\left({\tau\over 2M}\right) V_Q\left({\tau\over M}\right)
  V_{UV}\left({\tau\over 2M}\right) \right]^M \cdot
\nonumber \\ &&
V_{IR}\left({\tau\over 2}\right) ,
%\nonumber \\ &&
\label{intsch}\eea
where $M$ is a positive integer, and 
the effect of $V_Q, V_{UV}, V_{IR}$ on 
the system coordinates $\; \{ P,Q \}$
is given by
\bea && V_Q\left(\tau\right): \quad Q\rightarrow Q + \tau P \; ,
\nonumber \\ && 
V_{UV}\left(\tau\right): \quad P\rightarrow P - \tau \partial S_{UV} \; ,
\nonumber \\ && 
V_{IR}\left(\tau\right): \quad P\rightarrow P - \tau \partial S_{IR}\;\;. 
\eea
This integrator effectively contains two evolution time-scales, 
$\tau$ and $ \tau / M$. The choice of $M$ is a trade-off between 
the computational overhead from computing the force $\partial S_{UV}$ 
more frequently, 
and the gain from reducing the fluctuations of the HMC Hamiltonian 
at the end of the MD trajectory. In the case $M=1$ one gets an ordinary 
leap-frog integrator.

For testing the efficiency of our approach we performed numerical
simulations for three models. The first model is based on the action
(\ref{act0}). The other two models differ from each other 
by a different splitting (\ref{UVIR}) of the action (\ref{act1}):

\bi
\item {\bf Model A} 
\bea && 
 S_{UV} = S_G[U] , 
\nonumber \\ && 
 S_{IR} = S_{det}[U] + \phi^\dag (Q^\dag Q)^{-1} \phi \label{zrta}
\eea
\item {\bf Model B}
\bea && 
 S_{UV} = S_G[U] , 
\nonumber \\ && 
 S_{IR} = S_{det}[U] + 
 \phi^\dag W(Q^\dag Q)^{-1} W^\dag \phi \nonumber \\ &&
 + \chi^\dag (W^\dag W)^{-1} \chi \label{zrtb}
\eea
\item {\bf Model C}
%\bea && 
% S_{UV} = S_G[U] + S_{det}[U] +\chi^\dag (W^\dag W)^{-1} \chi , 
%\nonumber \\ && 
% S_{IR} = \phi^\dag W(Q^\dag Q)^{-1} W^\dag \phi 
%\label{zrt}\eea
\setlength{\arraycolsep}{0.1cm}
\bea 
 S_{UV} & = & S_G[U] + S_{det}[U] +\chi^\dag (W^\dag W)^{-1} \chi ,
\nonumber \\
 S_{IR} & = & \phi^\dag W(Q^\dag Q)^{-1} W^\dag \phi 
\label{zrt}\eea
\ei

Model A is just a standard HMC algorithm for which the original
splitting of the time-scale proposed by Sexton and Weingarten \cite{Sexton:nu} 
is applied. Model B is the modification proposed by Hasenbusch
\cite{Hasenbusch:2001ne}, which was numerically studied in 
Refs.~\cite{Hasenbusch:2002wi,Hasenbusch:2002ai}. Finally, 
model C is our proposal for introducing different time-scales for
the two parts of the pseudo-fermion action (\ref{act1}).

The splitting (\ref{zrt}) 
is motivated by the hypothesis that most of the 
high-frequency modes of the pseudo-fermion part of the 
action (\ref{act1}) are located in
$\chi^\dag (W^\dag W)^{-1} \chi$. We also put the 
clover determinant $S_{det}[U]$ on the ``ultraviolet'' time-scale because the 
force generated by it is computationally cheap. The computationally 
expensive term $\phi^\dag W(Q^\dag Q)^{-1} W^\dag \phi$ is put on the 
``infra-red'' time-scale.

\section{Simulation details and results}\label{ssimdet}

We tested the approach (\ref{zrt}) in production runs
on a $16^3\times 32$ lattice at
$\beta=5.29, \; \kappa=0.1355, \; c_{sw}=1.9192$ done by the 
QCDSF-UKQCD collaboration 
%Refs.~
\cite{Pleiter:2000qd},\cite{Stuben:2000wm},\cite{Booth:2001qp}.
These parameters
correspond to $ m_\pi / m_\rho \approx 0.7$.
The program was executed on the APEmille  
\cite{Bartoloni:2001he} at NIC Zeuthen.

For the fermion fields 
we use periodic (antiperiodic) 
boundary conditions in the spatial (time) directions.
A trajectory  
was composed of $N_{steps}$ consecutive steps
(\ref{intsch}), with the trajectory length equal to 1:
\be N_{steps}\cdot \tau = 1 \; . \ee
The linear equations appearing
in the calculation of the 
%inversion of the matrices $Q^\dag Q$ and $W^\dag W$, which is required for 
fermionic force and in updating $\phi ,\phi^\dag$ we
solve by the conjugate gradient algorithm.
In all cases the starting vector 
for the iterative solution
was the linear superposition of $N_{guess}$
solutions from the recent past \cite{Brower:1995vx}. 
In all our simulations we kept the value $N_{guess}=7$, which was empirically 
found to be close to the optimum.

We performed one run for model A, two runs for model B, and a few
runs for model C with different values for $\rho$ and $M$. All runs 
had a length of $300$ trajectories, which allowed us to get a 
reasonable estimate of the acceptance rates $P_{acc}$.
Our strategy was to try to keep the same acceptance 
rates for all runs by tuning the step-size $\tau$.

\begin{table*}[htb]
\caption{ Runs of 300 trajectories each on the $16^3\times 32$ 
lattice at $\beta=5.29, \kappa=0.1355, c_{sw}=1.9192$ for the models
of Eqs.~(\ref{zrta}), (\ref{zrtb}), (\ref{zrt}). Here
$\rho$ is the parameter in the operator (\ref{operrho}),
$M$ defines the second time-scale of the integration scheme (\ref{intsch}),
$N_{steps}=1/ \tau$ is the number of steps of which the trajectory
with length 1 was composed. $P_{acc}$ is the acceptance rate,
$N_Q$ and $N_W$ denote the average number of 
multiplications per trajectory by the matrices
$Q^\dag Q$ and $W^\dag W$, respectively.
$D_{gain}$ denotes the speed-up factor 
with respect to the standard HMC algorithm (model A).
\label{table1}}
\begin{center}
\begin{tabular}{|c|c|c|c|c|c|c|c|c|}
\hline
model & $\rho$ & $M$ & $N_{steps}$ &  $P_{acc}$ & $N_Q$ & $N_W$ & $N_Q+N_W$ & $D_{gain}$\\
\hline
{\bf A}  & 0   & 3 & 140 & 0.601 & 139492 & 0 & 139492 & 1 \\ 
\hline
{\bf B}  & 0.5 & 3 & 100 & 0.599 & 65951 & 5233 & 71184  & 1.95  \\ 
\hline
{\bf B}  & 0.2 & 3 & 70 & 0.664 & 47214  & 7378 & 54592 & 2.82 \\ 
\hline
{\bf C}  & 0.5 & 3 &  50 & 0.547 &  45160 & 7687 & 52847 & 2.40 \\ 
\hline
{\bf C}  & 0.2 & 3 &  40 & 0.663 &  32659 & 12373 & 45032 & 3.42 \\ 
\hline
{\bf C}  & 0.1 & 3 &  30 & 0.603 &  24932 & 15938 & 40870 & 3.42 \\ 
\hline
{\bf C}  & 0.07& 3 &  30 & 0.640 &  24512 & 20738 & 45250 & 3.28 \\ 
\hline
{\bf C}  & 0.1 & 5 &  30 & 0.733 &  24622 & 26235 & 50857 & 3.35  \\ 
\hline
\end{tabular}
\end{center}
\end{table*}

The main goal of this study was to compare the 
efficiencies of A, B, and C. These efficiencies 
are determined by the amount of CPU-time $t_{CPU}$ required  
for estimates of some 
observables with a given statistical error.
Since the computer time in simulations with dynamical fermions 
is mostly spent in the calculation of the pseudo-fermion force, the CPU-cost
is roughly proportional to
\be t_{CPU} \propto (N_Q + N_W)\cdot \tau_{int} \; .
\ee
Here $N_Q$ and $N_W$ denote the average number of 
multiplications by the matrices
$Q^\dag Q$ and $W^\dag W$, respectively, required for producing one 
MD trajectory, and $\tau_{int}$ is the integrated autocorrelation time for 
the observable under study. 
Unfortunately, our computer resources did not allow us to estimate
$\tau_{int}$ reliably. Therfore 
we base our investigation 
on the hypothesis that for fixed acceptance rates,
the autocorrelation times are the same for the 
different approaches and different
parameter sets considered in this paper, i.e.\  for all runs
\be \tau_{int} \propto\frac{1}{P_{acc}}\; . \ee
This hypothesis was confirmed by simulations of model B
on a $8^3\times 24$ lattice in Ref.~\cite{Hasenbusch:2002ai}.

Therefore, we measured the relative gain in computer time 
with respect to the standard HMC algorithm 
(model A) for the approach tested in the $i$-th
run by the following formula:
\be 
D_{gain}^{(i)} = \frac{N_Q^{(A)}}{N_Q^{(i)} + N_W^{(i)}}
\cdot 
\frac{P_{acc}^{(i)}}{P_{acc}^{(A)}} \; .
\ee
The larger the gain $D_{gain}^{(i)}$ for the $i$-th run, 
the less computer time is required
for estimating the observables by using the approach tested in 
that run.

We present our results in Table \ref{table1}. The following 
observations can be made:
\bi
\item
Putting the two contributions of the
pseudo-fermion part of the action (\ref{act1}) on
different time-scales of the integration scheme (model C)
gives some additional gain in computer time  
compared to the case, where the time-scale is 
the same for both parts (model B). 
A speed-up of $\;\approx 20\%$ is observed 
for $\rho=0.5$ and  $\rho=0.2$.
\item
In agreement with the studies in 
Refs.~\cite{Hasenbusch:2002wi,Hasenbusch:2002ai}, 
we observe that for fixed $M$ the performance
of the approach C is best for some optimal value
$\rho$, which in our case is likely to lie in the interval
$\rho\in [0.1,0.2]$ for $M=3$.
\item In one of the runs we increased the value of $M$
from 3 to 5, while $\rho=0.1$ was close to the optimal value.
We kept the same step-size $\tau$ for both runs. 
One sees that this change of $M$ 
increased the acceptance rate 
$P_{acc}$, but the gain in computer time $D_{gain}$ stayed
almost the same (or even slightly decreased)
due to the computational overhead coming from calculating the 
pseudo-fermion force $\partial S_{UV}$ more frequently.
\item
By using the approach C one achieves 
a speed-up of more than a factor three 
compared to the standard HMC algorithm A.
\ei

Our computational resources did not allow for a further resolution 
of the algorithmic performance in the space of the parameters $\rho, M$. 
Probably, the 
best improvement factor which we obtained, $D_{gain}=3.42$, can still be
slightly increased by further tuning of the parameters. However, we notice
that $D_{gain}$ 
seems to be quite stable for some range of the parameter $\rho$.
Therefore, 
we expect that not much tuning of the algorithm will be required in the
forthcoming production runs.

\section{Conclusions}\label{sconc}

In Refs.~\cite{Hasenbusch:2001ne,Hasenbusch:2001xh,Hasenbusch:2002wi,%
Hasenbusch:2002ai}
it was suggested to accelerate the HMC simulation of dynamical fermions
by splitting the fermion matrix into two factors with 
smaller condition numbers than that of the original matrix, and 
introducing pseudo-fermion fields for each of the factors. 
Inspired by the proposal of Ref.~\cite{Peardon:2002wb}, 
we tested the possibility to further speed up the simulations by putting 
each part of this new pseudo-fermion action on a separate time-scale. 
We have found that such a strategy gave a speed-up of $\;\approx 20\%$ 
in comparison to the case, where the time-scale was 
the same for both parts. 

In our simulations, which are a part of the production runs of the QCDSF 
collaboration, we have found a reduction of the numerical cost of more than a 
factor three compared to the standard HMC algorithm. One can speculate that
this reduction would have been even higher if we had not used  
the educated chronological guess \cite{Brower:1995vx}.
One can also expect that the gain
in computer time will grow as the quark mass decreases.

Further work in the direction of algorithmic improvement 
can be done by testing more complicated integration schemes 
than that of Eq.~(\ref{intsch}).
In Ref.~\cite{Hasenbusch:2002ai} it was shown that 
the splitting of the pseudo-fermion action (\ref{act1}) 
provided more computational gain
for the partially improved integration scheme suggested 
in Eq.~(6.4) of Ref.~\cite{Sexton:nu} 
than for the standard leap-frog scheme. It may be 
interesting to check the compatibility of that integration scheme 
(and higher order integration schemes)
with the multiple time-scales approach 
studied in our paper.

When going to smaller quark masses, 
a possibility might be to generalize 
the idea of Ref.~\cite{Hasenbusch:2001ne} by 
splitting the pseudo-fermion action into three or more parts
\cite{Hasenbusch:2002ai}. One can 
introduce different time-scales for each part of the pseudo-fermion action
in such an approach, profiting even more from the separation of high- and 
low-frequency modes.

\section*{ACKNOWLEDGEMENTS} 

The numerical calculations were performed on the APEmille at NIC (Zeuthen). 
%We thank the institution and personal for their support. 
We thank the operating staff for their support. 
TB acknowledges the  hospitality of NIC (Zeuthen) and
the Physics Department of Regensburg University.
%We are very grateful to R. Horsley and 
%P. Rakow 
%for useful discussions
%and help at the early stage of this project.
This work has been supported in part by the European
Community's Human Potential Program under contract HPRN-CT-2000-00145,
Hadrons/Lattice QCD.
TB acknowledges support from INTAS grant 00-00111,
Heisenberg-Landau program grant 2002-10, 
RFBR grants 02-01-00126 and 02-01-06064.

\end{document}